\documentclass{PoS}

\title{Dilepton measurements with CERES}

\ShortTitle{Dilepton measurements with CERES}

\def\rez{$^{\rm a}$}
\def\dub{$^{\rm b}$}
\def\gsi{$^{\rm c}$}
\def\fra{$^{\rm d}$}
\def\mpi{$^{\rm e}$}
\def\hei{$^{\rm f}$}
\def\wei{$^{\rm g}$}
\def\mue{$^{\rm h}$}
\def\sun{$^{\rm i}$}
\def\bnl{$^{\rm j}$}
\def\cer{$^{\rm k}$}

\author{Ana Mar\'{\i}n for the CERES Collaboration:
D.~Adamov\'{a}\rez,
G.~Agakichiev\dub,
A.~Andronic\gsi,
D.~Anto\'{n}czyk\fra,
H.~Appelsh\"{a}user\fra,
V.~Belaga\dub,
J.~Biel\v{c}\'{i}kov\'{a}\mpi$^,$\hei,
P.~Braun-Munzinger\gsi,
O.~Busch\gsi,
A.~Cherlin\wei,
S.~Damjanovi\'{c}\hei,
T.~Dietel\mue,
L.~Dietrich\hei,
A.~Drees\sun,
S.\thinspace I.~Esumi\hei,
K.~Filimonov\hei,
K.~Fomenko\dub,
Z.~Fraenkel\wei,
C.~Garabatos\gsi,
P.~Gl\"{a}ssel\hei,
G.~Hering\gsi,
J.~Holeczek\gsi,
M.~Kalisky\gsi,
S.~Kniege\fra,
V.~Kushpil\rez,
W.~Ludolphs\hei,
A.~Maas\gsi,
A.~Mar\'{\i}n\gsi,
J.~Milo\v{s}evi\'{c}\hei,
D.~Mi\'{s}kowiec\gsi,
R.~Ortega\hei,
Y.~Panebrattsev\dub,
O.~Petchenova\dub,
V.~Petr\'{a}\v{c}ek\hei,
M.~P\l{}osko\'{n}\fra,
S.~Radomski\hei,
J.~Rak\gsi,
I.~Ravinovich\wei ,
P.~Rehak\bnl,
H.~Sako\gsi,
W.~Schmitz\hei,
S.~Schuchmann\fra,
S.~Sedykh\gsi,
S.~Shimansky\dub,
R.~Soualah\hei,
J.~Stachel\hei,
M.~\v{S}umbera\rez ,
H.~Tilsner\hei,
I.~Tserruya\wei,
G.~Tsiledakis\gsi,
J.\thinspace P.~Wessels\mue ,
T.~Wienold\hei,
J.\thinspace P.~Wurm\mpi,
S.~Yurevich\gsi,
V.~Yurevich\dub \\
\llap{\rez}NPI ASCR, \v{R}e\v{z}, Czech Republic\\
\llap{\dub}JINR Dubna, Russia\\
\llap{\gsi}GSI Darmstadt, Germany\\
\llap{\fra}Frankfurt University, Germany\\
\llap{\mpi}MPI, Heidelberg, Germany\\
\llap{\hei}Heidelberg University, Germany\\
\llap{\wei}Weizmann Institute, Rehovot, Israel\\
\llap{\mue}M\"{u}nster University, Germany\\
\llap{\sun}SUNY Stony Brook, U.S.A.\\
\llap{\bnl}BNL, Upton, U.S.A.\\
\llap{\cer}CERN, Geneva, Switzerland\\

        E-mail: \email{a.marin@gsi.de}}

\abstract{We report on dilepton measurements for central Pb on Au collisions at the top 
CERN SPS energy with the upgraded CERES experiment. The dilepton mass spectrum 
of 2000 data with improved mass resolution shows an enhancement over the expectation 
from hadron decays that is well described by a model including a strong broadening of 
the $\rho$ spectral function. The measured excess yield excludes the dropping mass scenario. \\
We also report on the $\phi$ meson measured simultaneously both in the $K^+K^-$ and 
in the dilepton decay channel for the first time in high energy heavy-ion collisions. 
An excellent agreement is found between the rapidity densities and the shape 
of the measured tranverse momentum spectrum. The data rule out a possible enhancement 
of the $\phi$ yield in the leptonic over hadronic channel by a factor larger than 
1.6 at 95\% CL. CERES results are in agreement with NA49 results.}

\FullConference{Critical Point and Onset of Deconfinement   -  4th International Workshop\\
                 July 9 - 13, 2007\\
                 Darmstadt, Germany}

\begin{document}
\section{Introduction}
The main goal of ultra-relativistic heavy-ion collisions is the study of very hot 
and dense nuclear matter. This kind of matter is believed to have existed shortly after 
the {\it Big Bang}. Lattice QCD calculations \cite{lattice} have predicted a transition 
from ordinary hadronic matter to a plasma of quarks and gluons at high energy densities. 
At the same time Chiral symmetry is restored. Dileptons that carry information of the 
entire fireball evolution are very suitable probes because of their negligible final state 
interactions.

Medium modifications of the $\phi$ meson properties (mass and width) or of the kaons
that might result in a change  of the branching fraction of $\phi$ decaying to K$^+$K$^-$ or 
to $e^+e^-$ when $\phi$ decays in medium, may be related to the expected chiral phase 
transition \cite{koc2}. However as the $\phi$ lifetime (44 fm/c) is longer than the 
expected lifetime of the coupled collision system, only a fraction may decay in the hot 
fireball. Final state interactions of kaons from $\phi$ meson decays may lower
the measured branching ratio into the kaon channel \cite{joh}. Indeed, the NA50 experiment
measuring the $\phi$ meson in the $\mu^+\mu^-$ decay channel and NA49 measuring in 
the K$^+$K$^-$ channel, obtain yields that differ by factors between 2 and 4 in the common
$m_t$ range \cite{puzzle}. Further, the $m_t$ spectra exhibit a different inverse
slope parameter, 305~$\pm$~15 MeV in NA49 and 218~$\pm$~6 MeV in NA50,
fitted in their $m_t$ acceptance regions.

The CERES experiment at the CERN SPS is an experiment dedicated to the study of
low mass dilepton pairs. CERES has measured an enhanced dilepton production in the
invariant mass region $m_{e^+e^-}>$~0.2~GeV/c$^2$ in S+Au at 200~AGeV \cite{exp} and 
in Pb+Au at 158~AGeV \cite{eleletter9596}. The enhancement is 
absent in p-induced reactions \cite{pp}. Pion annihilation has been taken into
account as an additional mechanism for e$^+$e$^-$ production but the 
experimental spectra cannot be explained without introducing medium modifications of 
vector mesons, particularly of the $\rho$. The two main theoretical choices for this 
modification are the dropping mass scenario \cite{theo2} and the broadening scenario \cite{theo1}, 
but the experimental precision did not allow to distinguish between them.

In order to further investigate the enhancement and possibly discriminate between the different
theoretical approaches, the CERES spectrometer was upgraded during 1998 by the addition of 
a Time Projection Chamber (TPC) with radial electric drift field \cite{up2,up1,qm99,qm04,qm05,tpcnim} 
which improves the mass resolution and the electron identification. 
In order to investigate the role of baryons, during 1999, CERES took data with the TPC 
(although with partial readout) at a bombarding energy of 40~AGeV \cite{eleletter1999} 
where the baryon density is approximately 30\% larger than at 158~AGeV \cite{na49protons}.
During the year 2000 CERES took a large data sample consisting of
30$\cdot10^6$ and 3$\cdot10^6$ events of Pb on Au collisions at 158 AGeV triggered on the
7\% and 20\% most central collisions \cite{qm05}, respectively. 
Moreover, the upgrade of the CERES experiment makes possible for the first time 
in high energy heavy-ion collisions to study simultaneously the leptonic and the 
charged kaon decay modes of the $\phi$ meson at the SPS, thus shedding light 
onto the $\phi$ puzzle.

In this paper, we present an overview of the most relevant dilepton results of the CERES 
experiment since the upgrade with the TPC. All these results have been previously published 
in \cite{eleletter1999,eleletter2000,philetter}.

\section{Experimental setup}

\begin{figure}
\includegraphics[width=0.6\textwidth,angle=-90]{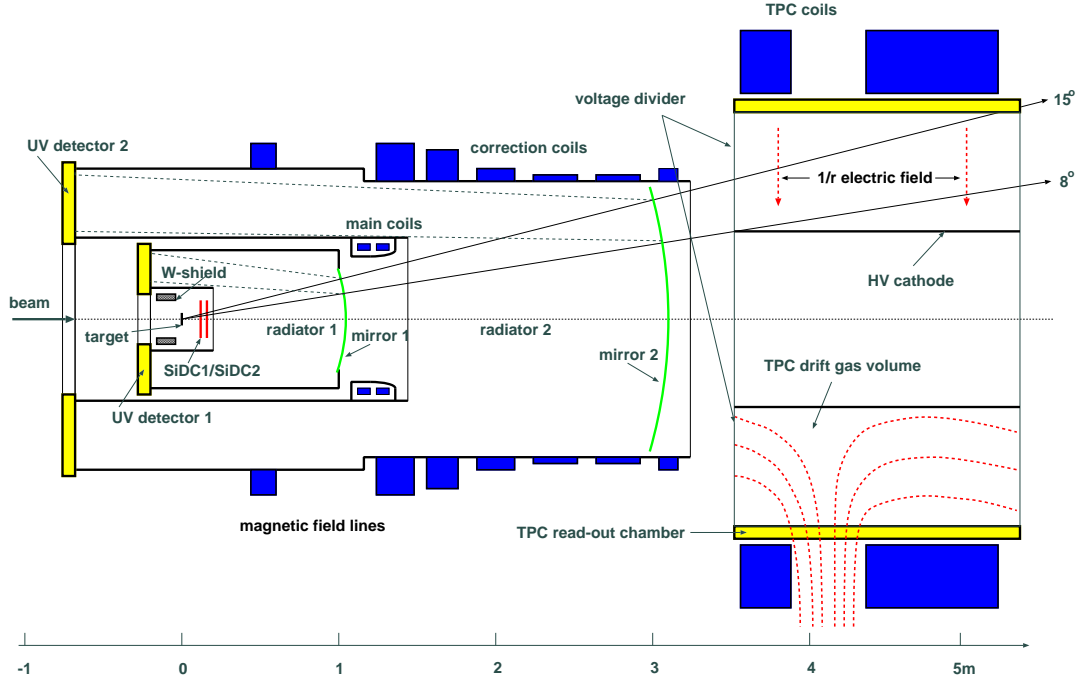}
\caption{Cross section through the upgraded setup of the CERES
spectrometer. The apparatus has a cylindrical symmetry.}
\label{fig:setup}
\end{figure}
The CERES experiment (Fig.~\ref{fig:setup}) is optimized to measure 
low mass electron pairs close to mid-rapidity (2.1$<\eta<$2.6) with 
full azimuthal coverage. A vertex telescope, composed of two Silicon 
Drift Detectors (SDD) positioned at 10.4 cm and 14.3 cm downstream of a 
segmented Au target, provides a precise vertex reconstruction, 
angle measurement for charged particles and rejection of close pairs
from $\gamma$ conversions and $\pi^0$ Dalitz decays. Two Ring Imaging 
CHerenkov (RICH) detectors, operated at a high threshold ($\gamma_{th}$=32),
 are used for electron identification in a large hadronic background. 
In the configuration with the TPC, the magnetic field between the
two RICH detectors is switched off, allowing to use them in a combined 
mode resulting in an increased electron efficiency. The new radial-drift TPC, 
positioned downstream of the original spectrometer, has an active length of 2 m 
and a diameter of 2.6~m. A gas mixture of Ne (80\%) and CO$_2$ (20\%) 
is used. It is operated inside a magnetic field (indicated 
by the dashed field lines inside the TPC in Fig.~\ref{fig:setup}) 
with a maximal radial component of 0.5 T and provides up to 20
space points for each charged particle track. This 
is sufficient for the momentum determination with a resolution 
$\Delta p/p\sim((2\%)^2+(1\%\cdot p$(GeV/c))$^2)^{1/2}$
 and for additional electron identification by using the d$E$/d$x$ signal in the TPC.

\section{Di-electron analysis}
Charged particles from the target are reconstructed by matching track 
segments in the SDD and in the TPC using a momentum-dependent matching window. 
Tracks in the TPC are required to contain more than 12 hits out of a maximum 
20 possible, to ensure good momentum resolution. Cherenkov rings with asymptotic 
radii are identified in the RICH using a Hough transformation (see sect. 3.2.4 
in ref. \cite{eleletter9596}). The TPC electron selection is done based on the 
d$E$/d$x$ signal and its resolution. To select electrons among all charged 
hadrons both a TPC d$E$/dx signal in the electron region and a matching to a 
RICH ring are required. The combined pion rejection factor varies 
from 4$\times$10$^4$ to 1.8$\times$10$^4$ for momenta between 1~GeV/c 
and 2.5~GeV/c for total electron efficiencies of 68\% and 66\% \cite{serguei}, 
respectively, with the quality cuts applied in the analysis \cite{busch,serguei,alex}.

The main difficulties of the electron analysis are the low probability of electromagnetic 
decays and the large amount of combinatorial background from $\gamma$ conversions 
and Dalitz decays. Very good electron identification is not enough. Electron pairs 
from $\gamma$ conversions and $\pi^0$ Dalitz decays, characterized by their small 
opening angle and low momentum, need to be identified by their topology and removed 
from the sample in order to reduce the combinatorial background. As the detectors have a
finite two-track resolution the most effective way of rejecting
conversions in the target and close Dalitz pairs is
by rejecting tracks with an energy loss signal significatively larger than 
the minimum ionizing energy loss signal in both SDD's where all
hit amplitudes have been re-summed around each SDD track
segment. Low-amplitude tails in the SDD d$E$/dx distribution are also removed. 
Late conversions (mostly in SDD1) are removed by a cut in the distance to
another TPC track of opposite sign and with electron d$E$/dx. 
To reduce $\pi^0$ Dalitz decay contributions only electron tracks 
with no opposite charge electron track within 35 mrad are taken for
analysis. Finally, only electron tracks in the geometrical 
acceptance 0.141~rad~$< \theta< $~0.243 rad and with
a $p_t>$~0.2~GeV/c are selected. 

\begin{figure}[hbt]
\begin{center}
\includegraphics[width=0.5\textwidth]{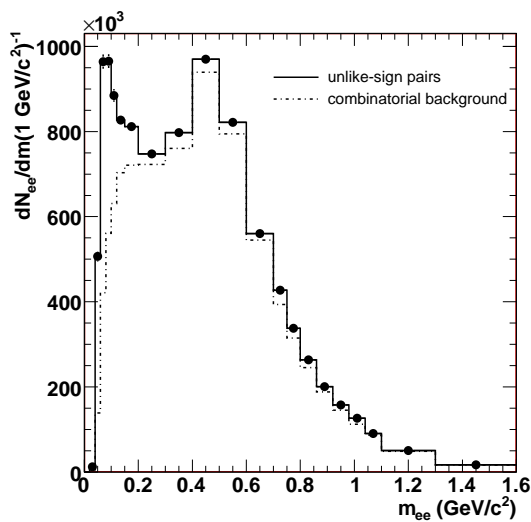}
\label{fig:fullrejectmix}
\caption{ Invariant-mass distribution of unlike-sign pairs (histogram), 
and mixed-event background (dashed line) normalized
to like-sign pairs background after full rejection and corrected 
for pair reconstruction efficiency event by event.}
\label{fig:sigback}
\end{center}
\end{figure}

The invariant-mass distributions of unlike-sign pairs and of the combinatorial 
background after full rejection and corrected for pair reconstruction efficiency 
event by event are shown in Fig.~\ref{fig:sigback}. 
In order to reduce the statistical errors an unlike-sign combinatorial background 
using the mixed-event technique has been evaluated. The mixed-event background is
normalized to the like-sign pair background in the mass region m$_{e^+e^-}>$~0.2~GeV/c$^2$.
For masses below 0.2~GeV/c$^2$, the like-sign background is used. 
The physics signal is obtained by subtracting the like-sign pairs or
the mixed event background from the unlike-sign pairs.

\section{Low-mass e$^+$e$^-$ results}

\begin{figure}
\includegraphics[width=0.5\textwidth]{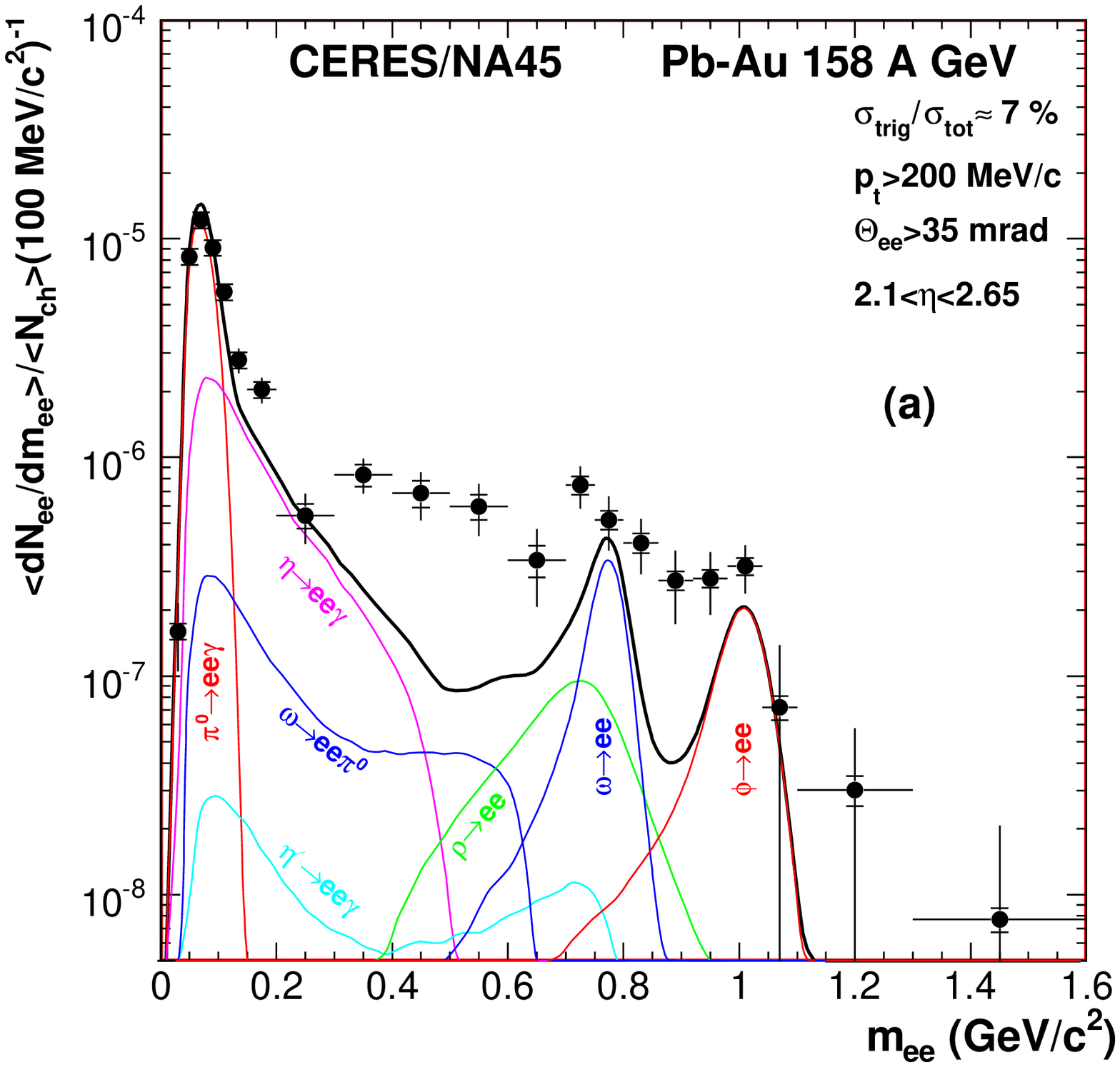}
\includegraphics[width=0.5\textwidth]{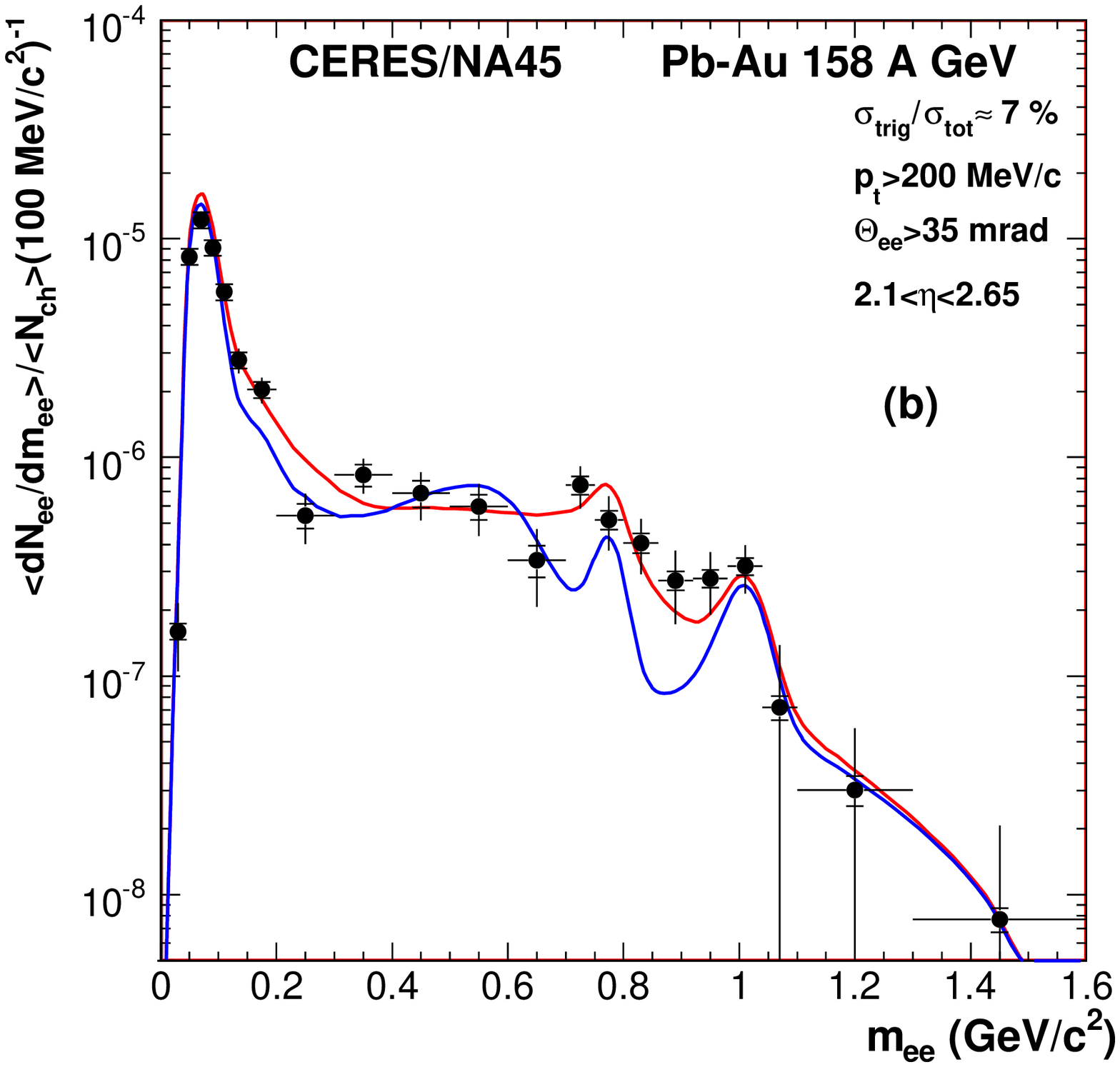}

\caption{(a) Invariant-mass spectrum of $e^+e^-$-pairs
compared to the expectation from the hadron decay 
cocktail. (b) The expectations from model calculations assuming
a dropping of the $\rho$ meson mass (blue) or a spread of the 
$\rho$ width in the medium (red) are also shown.}
\label{fig:fullreject}
\end{figure}

The e$^+$e$^-$ invariant-mass spectrum corrected event-by-event for the 
pair reconstruction efficiency and normalized to the number of events and 
to the average charged particle multiplicity $<N_{ch}>$ in the 
acceptance \cite{qm05} is shown in Fig.~\ref{fig:fullreject}(a) \cite{eleletter2000} 
compared to the expectations from the hadronic decay cocktail. The hadron decay cocktail
\cite{cocktail} has been folded with the experimental momentum resolution 
and energy loss due to bremsstrahlung. Acceptance, opening-angle, and 
transverse-momentum cuts are applied. An excess of pairs for 
m$_{e^+e^-}>$ 0.2~GeV/$c^2$ is clearly visible. The number of pairs 
in the Dalitz region (m$_{e^+e^-}<$~0.2~GeV/c$^2$) is 6114~$\pm$~176 with a signal 
to background ratio (S/B) of 1/2. The number of open pairs 
(m$_{e^+e^-}>$~0.2~GeV/c$^2$) is 3115~$\pm$~376 with a S/B of 1/22. 
The enhancement factor for 0.2~GeV/c$^2$$<$ m$_{e^+e^-}$$<$1.1~GeV/c$^2$ 
compared to the hadron decay cocktail 
is 2.56 $\pm$ 0.22 (stat) $\pm$ 0.31 (syst) $\pm$ 0.83 (decays) 
\footnote[1]{The change on the branching ratio of the proccess $\omega\rightarrow \pi^0e^+e^-$ 
and the improvement in its error in PDG 2006 have not been considered yet.}.
The total data systematic uncertainty of 12\% includes the combinatorial
background subtraction and the electron efficiency correction.
The systematic error of the charged particle multiplicity determination of 12\% has been included
on the systematic error of the cocktail.

The experimental results are compared to theoretical
models \ref{fig:fullreject}(b) based on hadronic decays and $\pi^+\pi^-$ annihilation. 
The $\rho$-propagator is treated \cite{rapp} in 3 ways: vacuum $\rho$ (not shown), 
modifications following Brown-Rho scaling \cite{theo2}, and modifications 
via $\rho$-hadron interactions \cite{theo1}. 
In the region between the $\omega$ and the $\phi$, measured now with  
better resolution, the data favor the many-body approach over Brown-Rho scaling.

\section{The in-medium $\rho$}

\begin{figure}
\begin{center}
\includegraphics[width=0.5\textwidth]{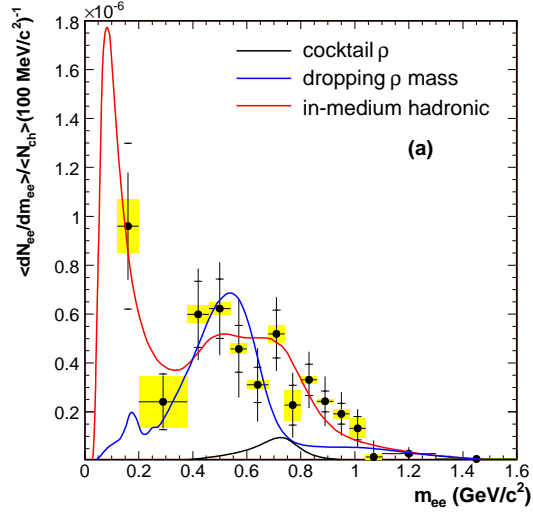}
\end{center}
\caption{(a) Dilepton yield after subtraction of the hadronic cocktail 
(without the $\rho$). The systematic errors of the data (horizontal ticks) 
and  the systematic uncertainty of the subtracted cocktail (shaded boxes) are
also shown. The data are compared to expectations of the models assuming a shift 
of the in-medium $\rho$ mass (blue) and a broadening scenario (red). 
%(b) The excess yield is compared to the expectations of a 
%broadening scenario with (red) and without baryon effects (blue). 
}
\label{fig:subspec}
\end{figure}

In order to further investigate the excess dilepton yield the contribution from the 
hadronic cocktail (except the $\rho$) is subtracted from the data and from the 
calculations (Fig. \ref{fig:subspec}a). No rescaling of theory to data is used.
It can be observed that the contribution of the 
cocktail $\rho$ is totally negligible. The in-medium $\rho$ contribution dominates
by more than one order of magnitude (enhancement factor of 16.4$\pm$1.8 
for 0.12 GeV/c$^2<$m$_{e^+e^-}<$ 1.1 GeV/c$^2$) in central Pb on Au collisions. The shape and 
the yield are well described by the in-medium hadronic contribution. A detailed study 
of the spectrum width including systematic and statistical errors excludes the 
dropping $\rho$ mass scenario \cite{eleletter2000}.

%The larger e$^+$e$^-$ enhancement measured at 40AGeV (with 1.8$\sigma$ significance) 
%\cite{eleletter1999} with respect to the one measured at 158 AGeV indicated that the baryon 
%density is more important than the temperature for the in-medium modifications of the $\rho$ meson.
%The role of baryons can also be address in the 2000 data by comparing the excess yield to a calculation
%of the broadening scenario with and without baryon-induced interactions (Fig. \ref{fig:subspec}b).
%As one can observe the calculation without baryon-induced interactions underestimates the data
%for masses below 0.5 GeV/c$^2$ while the calculation including baryon-induced interactions
%describes the data better.

\section{The $\phi$ meson}

The $\phi$ meson yield in the e$^+$e$^-$ channel is determined by integrating the
invariant mass spectra in the mass region between 0.9 and 1.1 GeV/c$^2$ 
in three transverse momentum bins \cite{philetter}. The integrated yield in the mass range 
between 0.9 and 1.1 GeV/c$^2$ is 229$\pm$53 with a signal to background 
ratio of 1/12 and needs to be corrected for acceptance, reconstruction efficiency 
and physics background under the $\phi$ peak.
The $\rho$ meson could extend into this mass range if 
its spectral function is modified in the medium. 
Dileptons from the QGP phase also contribute to the physics background in this mass range.
The sum of these two contributions is estimated to be 35\% 
of the total yield in this mass region by inspecting theoretical models
that include in-medium spreading of the $\rho$ width due to 2$\pi$ processes and 
the dilepton yield from the QGP phase \cite{theo1}. 
%If processes involving 4 or 6 pions 
%would be included in \cite{theo1} the physics background could be larger.
%Using another model \cite{theo3} that also describes the CERES data gives a physics background  
%contribution of 37\%. 
The measured $\phi$ yield has been scaled by this factor to correct for the physics background.
The charm contribution that is smaller than 3\% \cite{charm} has been neglected.

As mentioned before, CERES can also study the $\phi$ meson in the charged 
kaon (K$^+$K$^-$) decay mode. In order to do so, all charged
particles get assigned the kaon mass (no particle identification is used). 
Only a conservative upper cut in the d$E$/d$x$ signal (corresponding to 90\% of the 
Fermi plateau value) for momenta between 1.25 GeV/c and 4 GeV/c, suppressing 83\% of the 
electrons, is applied to enhance the kaon content of the sample.
Tracks in the geometrical acceptance 0.13~rad
$<\theta<$~0.24~rad with a transverse momentum $p_t$ larger than 0.25~GeV/c are selected.
To reduce the contamination from other particle species, cuts in the Podolanski-Armenteros 
parameter \cite{armen} and in the opening angle between the kaons are applied.
The $\phi$ meson in the kaon decay mode is studied in the rapidity
interval 2.0~$<y^\phi<$~2.4 for $p_t^\phi>0.75$~GeV/c.

\begin{figure}
\begin{center}
\includegraphics[width=0.5\textwidth]{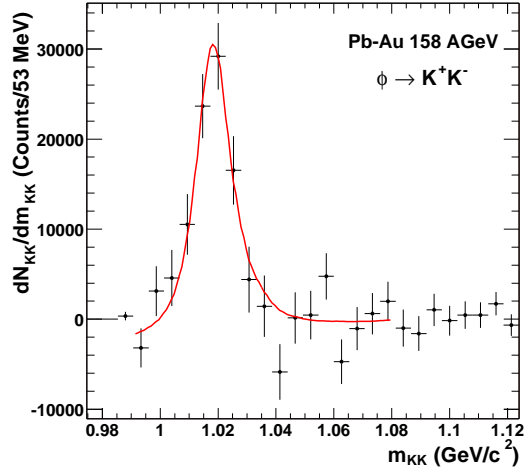}
\caption{Invariant-mass spectra of $K^+K^-$ pairs after background
subtraction for the ($p_t^\phi$,$y^\phi$) bin 1.5 GeV/c $<$ $p_t^\phi$ $<$ 1.75 GeV/c
and 2.2 $<$ $y^\phi$ $<$ 2.4.}
\label{fig:phi_ka}
\end{center}
\end{figure}

The invariant mass distributions of K$^+$K$^-$ pairs were
accumulated in ($p_t^\phi$,$y^\phi$) bins to obtain the $\phi$ 
transverse momentum spectrum. The combinatorial background invariant 
mass distributions are calculated using the mixed-event technique for 
each ($p_t^\phi$,$y^\phi$) bin. An example of invariant mass spectrum after background subtraction 
is presented in Fig.~\ref{fig:phi_ka}.
The yield of the $\phi$ mesons is determined by fitting a relativistic 
Breit-Wigner distribution with parameters taken from the Particle Data Group compilation \cite{pdg}
(convoluted with the experimental resolution function obtained by a Monte Carlo simulation)
superimposed on a linear background to account for a residual background in the low $p_t^\phi$ bins, 
to the measured line shape.
 The signal to background ratios vary from 1/2000 to 1/180 with increasing $p^\phi_t$.
The signal is integrated in the mass range between 1.0 GeV/c$^2$ and 1.05 GeV/c$^2$.
The resulting $\phi$ yields were corrected for acceptance and efficiency.

\begin{figure}[hbt]
\includegraphics[width=0.5\textwidth]{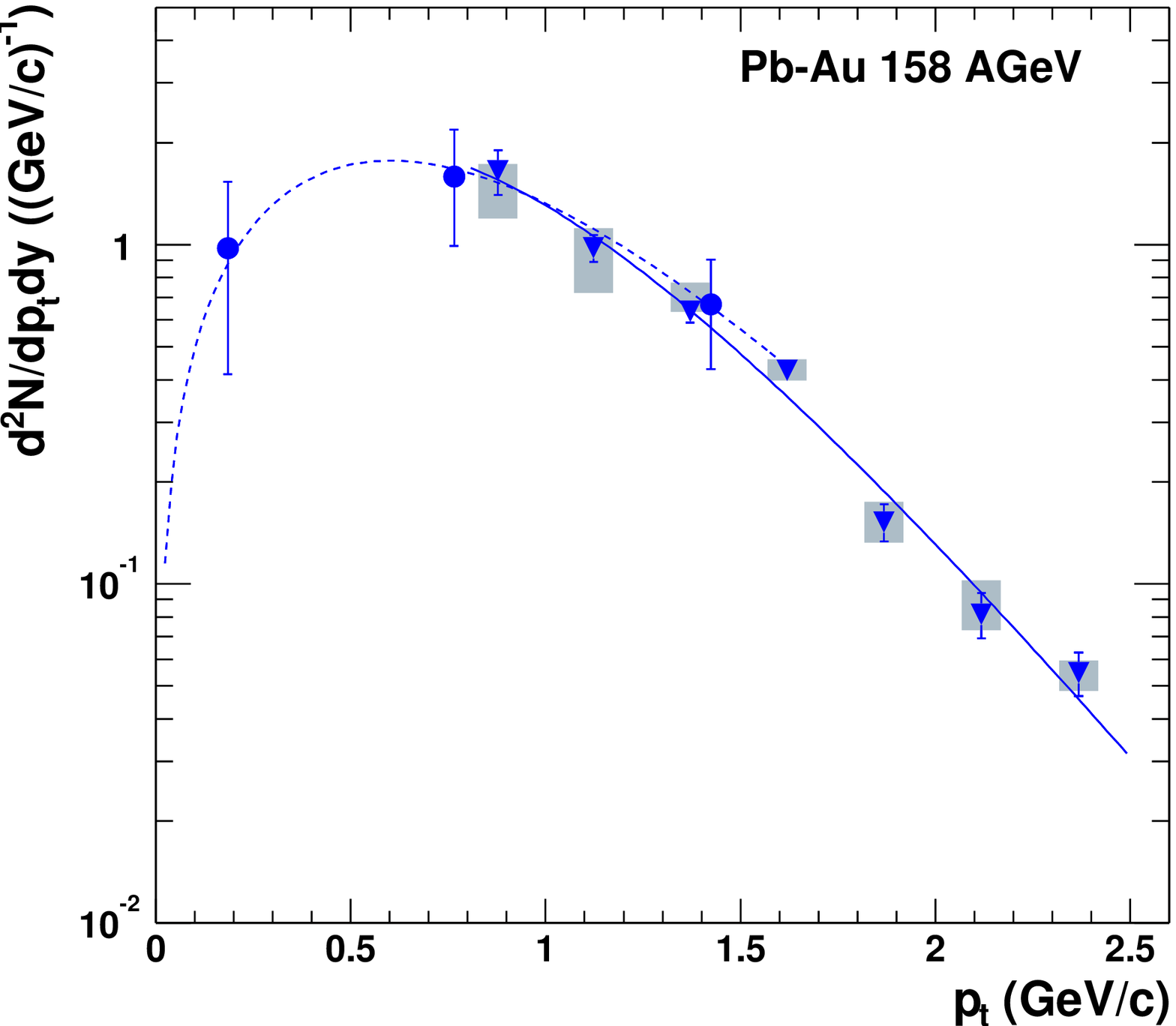}
\includegraphics[width=0.42\textwidth]{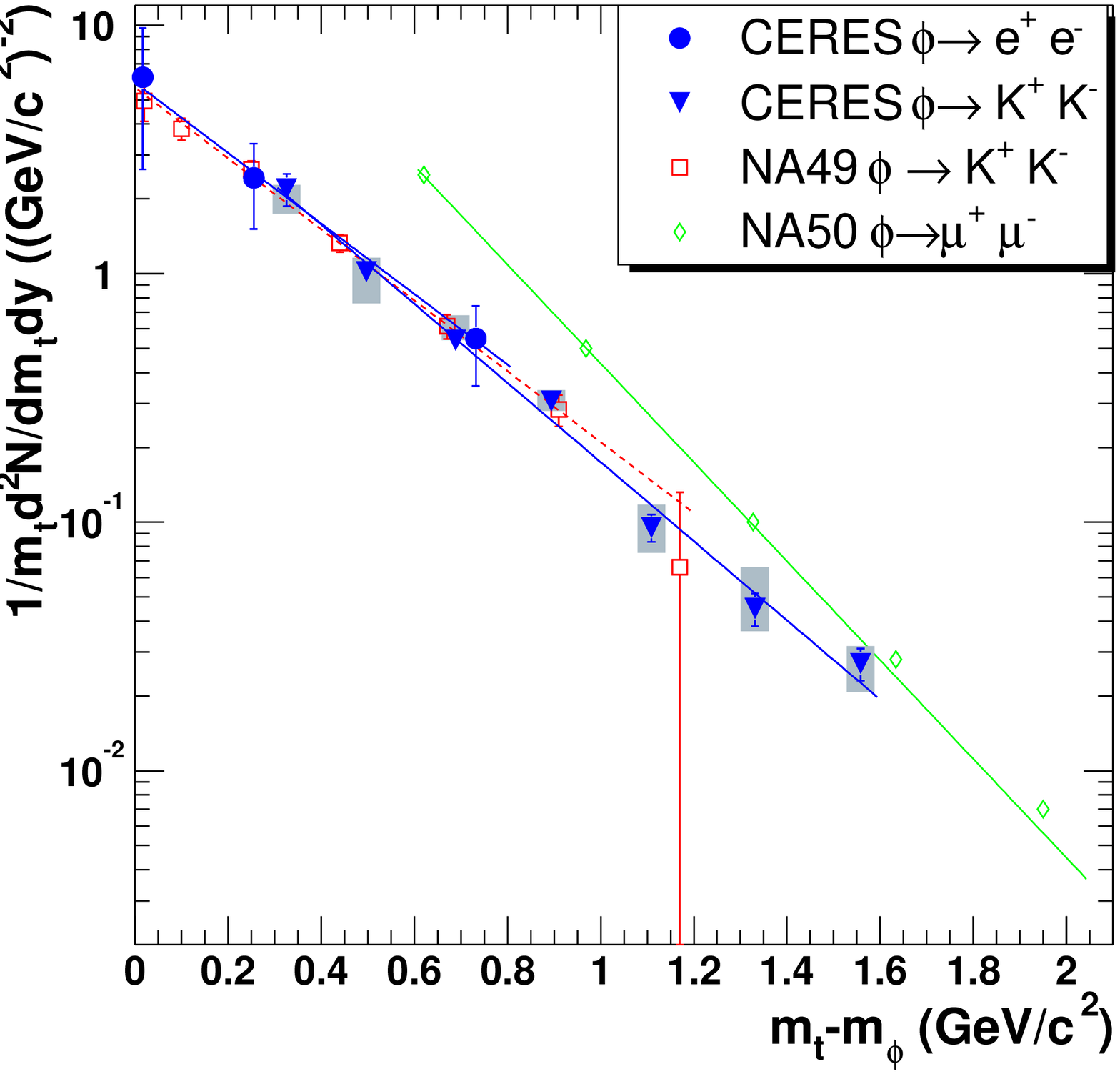}
\caption{Left: Transverse momentum spectrum of $\phi$ mesons corrected for
acceptance and efficiency reconstructed in the e$^+$e$^-$ decay 
mode (circles) and in the K$^+$ K$^-$ decay channels (triangles). 
Systematic errors in the kaon decay channel are shown as boxes. The systematic errors in the dilepton analysis
(not shown) are $\pm$16\%.
Right: Transverse mass distribution of $\phi$ mesons measured in the
charged kaon (triangles) and in the dilepton (circles) decay mode 
after scaling (see text) compared to the results from NA49 (squares) and NA50 (diamonds).} 

\label{fig:mtcomp}
\vspace*{0.5cm}
\end{figure}

The efficiency- and acceptance-corrected $\phi$ meson yield is shown in 
Fig.~\ref{fig:mtcomp} (left) for both decay modes as a function of transverse momentum. 
The inverse slope parameter of $T$~=~273~$\pm$ 9(stat)~$\pm$ 10(sys) MeV and a rapidity density 
$dN/dy$ of 2.05~$\pm$~0.14(stat) $\pm$~0.25(sys) in the K$^+$K$^-$ decay mode 
and $T$~=~306~$\pm$~82(stat)~$\pm$~40(syst) MeV and 
$dN/dy$~= 2.04~$\pm$~0.49(stat)~$\pm$~{0.32}(syst) in the dilepton decay 
mode are in good agreement within the errors.
A $\phi$ meson yield in the e$^+$e$^-$ decay mode larger 
than 1.6 times the yield on the K$^+$K$^-$ decay mode is excluded at 95\% CL 
(statistical and systematic errors in both decay channels added in quadrature).

Moreover, the CERES results can be compared to the existing Pb-Pb systematics \cite{puzzle} 
after accounting for the different measurement conditions.
 The NA49 measurement was done at 4\% centrality and covered a rapidity range from 3 to 3.8
units \cite{Friese}. 
A global scaling factor of $1.17 \pm 0.12$ obtained experimentally 
is applied to the combined CERES data of Fig.~\ref{fig:mtcomp} (left) to make the comparison 
to the systematics of \cite{puzzle}.
In Fig.~\ref{fig:mtcomp} (right) the scaled CERES $\phi$ transverse mass spectrum 
is plotted together with the NA49 and NA50 data.
The $\phi$ meson yields agree within the errors with the NA49 results. So does 
the yield in the K$^+$K$^-$ extrapolated down to $p_t$=0 using the measured inverse 
slope parameter.
On the other hand, CERES data in the K$^+$K$^-$ channel do not agree with NA50 results in
the common $p_t$ region. This experiment however measures the leptonic channel. 
The extrapolation of NA50 results down to the region where CERES measures the 
dilepton channel does not agree either. As stated above, in the CERES measurement the 
two decay modes agree. Possible differences of maximum 40-50\% as expected by models including 
only rescattering of the kaons \cite{joh} or of maximum 70\% at the 
lowest $p_t$ ($p_t < 0.3$ GeV/c) expected by 
models including medium modifications of the $\phi$ mesons and kaons like the 
AMPT model \cite{pal02} cannot be ruled out by the CERES results. 

\section{Conclusions}
To conclude, CERES has measured the low invariant mass spectrum of dilepton pairs with improved mass 
resolution. An enhancement of 2.56 $\pm$ 0.22 (stat) $\pm$ 0.43 (syst) $\pm$ 0.76 (decays) over 
the expectations of the hadronic decay cocktail is measured.
The excess dilepton yield clearly favors models including a strong broadening of the
$\rho$ spectral function and it rules out models including a mass shift of the $\rho$ meson. 

At 40~AGeV, the observed pair yield is enhanced over the expectation from neutral meson decays by a 
factor of 5.9 $\pm$ 1.5(stat) $\pm$ 1.2(syst) $\pm$ 1.8(decays). Compared to 
158~AGeV it is somewhat larger, with a significance of 1.8 $\sigma$. The fact that the enhancement 
is somewhat larger and that the only quantity rising is the baryon density while the temperature 
decreases shows the importance of the baryon density for the in medium modifications of the $\rho$ meson.

Moreover, $\phi$ meson production has been measured simultaneously in both decay channels 
for the first time in relativistic heavy-ion collisions. The yield and the inverse slope parameter in both 
decay modes agree within the errors. 
Our results are in agreement with the results from NA49 measured in the kaon channel. 
A yield in the e$^+$e$^-$ decay mode larger 
than 1.6 times the yield on the K$^+$K$^-$ is excluded at 95\% CL, therefore the large discrepancy 
observed previously is not observed in the CERES data.
The theoretical predictions in \cite{joh,pal02} are consistent with our data.

\section{Acknowledgments}
The CERES collaboration acknowledges the good performance of the CERN
PS and SPS accelerators as well as the support from the EST
division. We would like to thank R.~Campagnolo, L.~Musa, A.~Przybyla, W.~Seipp 
and B.~Windelband for their contribution during construction and commissioning 
of the TPC and during data taking. We are grateful for excellent support by the 
CERN IT division for the central data recording and data proccesing. This work was 
supported by GSI, Darmstadt, the German BMBF, the German VH-VI 146, the US DoE, 
the Israeli Science Foundation, and the MINERVA Foundation.

\end{document}